\documentclass[conference,onecolumn]{IEEEtran}

\usepackage{amsmath,amssymb,amsfonts}
\usepackage{graphicx}
\usepackage{textcomp}
\usepackage{xcolor}

\usepackage{subcaption}
\usepackage{enumitem}

\usepackage{verbatim} 
\usepackage{multicol} 

\usepackage[switch]{lineno}

\usepackage[hidelinks]{hyperref}
\usepackage{comment}

\usepackage{algorithm}
\usepackage{listings}
\usepackage{algpseudocode}

\usepackage{url}
\usepackage{bm}

\usepackage{tikz}

\def\BibTeX{{\rm B\kern-.05em{\sc i\kern-.025em b}\kern-.08em
    T\kern-.1667em\lower.7ex\hbox{E}\kern-.125emX}} 

\graphicspath{{figs}} 

\usepackage{multirow}   

\def\BibTeX{{\rm B\kern-.05em{\sc i\kern-.025em b}\kern-.08em
    T\kern-.1667em\lower.7ex\hbox{E}\kern-.125emX}}

\newcommand{\bfB}{\mathbf{B}}
 
\newcommand{\bB}{\mathbf{B}}

\newcommand{\bJ}{\mathbf{J}}

\newcommand{\bv}{\mathbf{v}}

\newcommand{\br}{\mathbf{r}}

\newcommand\copyrighttext{%
  \footnotesize \textcopyright \the\year{} IEEE. Personal use of this material is permitted. Permission from IEEE must be obtained for all other uses, including reprinting/republishing this material for advertising or promotional purposes, collecting new collected works for resale or redistribution to servers or lists, or reuse of any copyrighted component of this work in other works.}

\newcommand\copyrightnotice{%
\begin{tikzpicture}[remember picture,overlay]
\node[anchor=south,yshift=10pt] at (current page.south) {\fbox{\parbox{\dimexpr0.75\textwidth-\fboxsep-\fboxrule\relax}{\copyrighttext}}};
\end{tikzpicture}%
}

\begin{document}

\bstctlcite{IEEEexample:BSTcontrol}

\title{Accelerating the Particle-In-Cell code ECsim with OpenACC
}

\author{\IEEEauthorblockN{
Elisabetta Boella\IEEEauthorrefmark{1}\IEEEauthorrefmark{5},
Nitin Shukla\IEEEauthorrefmark{2}\IEEEauthorrefmark{6},
Filippo Spiga\IEEEauthorrefmark{3},
Mozhgan Kabiri Chimeh\IEEEauthorrefmark{3},
Matt Bettencourt\IEEEauthorrefmark{3},
Maria Elena Innocenti\IEEEauthorrefmark{4}
}\\
\IEEEauthorrefmark{1}\textit{E4 Computer Engineering SpA}, Scandiano, Italy\\
\IEEEauthorrefmark{2}\textit{CINECA}, Casalecchio di Reno, Italy\\
\IEEEauthorrefmark{3}\textit{NVIDIA Corporation}, Santa Clara, United States\\
\IEEEauthorrefmark{4}\textit{Ruhr-Universit\"at}, Bochum, Germany\\

\IEEEauthorrefmark{5}\textit{elisabetta.boella@e4company.com}, 
\IEEEauthorrefmark{6}\textit{n.shukla@cineca.it}\\\vspace*{-1cm}
}


\maketitle
\copyrightnotice

\begin{abstract}
The Particle-In-Cell (PIC) method is a computational technique widely used in plasma physics to model plasmas at the kinetic level. In this work, we present our effort to prepare the semi-implicit energy-conserving PIC code ECsim for exascale architectures. To achieve this, we adopted a pragma-based acceleration strategy using OpenACC, which enables high performance while requiring minimal code restructuring. On the pre-exascale Leonardo system, the accelerated code achieves a $5 \times$ speedup and a $3 \times$ reduction in energy consumption compared to the CPU reference code. Performance comparisons across multiple NVIDIA GPU generations show substantial benefits from the GH200 unified memory architecture. Finally, strong and weak scaling tests on Leonardo demonstrate efficiency of $70 \%$ and $78 \%$ up to 64 and 1024 GPUs, respectively. 
\end{abstract} 
\begin{IEEEkeywords}
Particle-In-Cell, Plasma Physics, GPU acceleration, OpenACC, performance optimisation, energy efficiency.
\end{IEEEkeywords}

\section{Introduction}

The Particle-In-Cell (PIC) method is a specialised par\-ti\-cle-mesh technique widely used in plasma physics to mod\-el kinetic effects and complex plasma interactions. It combines Lagrangian and Eulerian approaches to accurately capture plasma microphysics. Specifically, plasmas are represented through a statistical distribution of charged particles, with computational macroparticles sampling the behavior of positive and negative charges. These macroparticles interact via the electromagnetic fields they generate, which are computed by solving Maxwell's equations on a fixed grid. The charge and current densities required for field calculations are obtained by interpolating the discrete particle properties onto the grid. In turn, the fields are used to update the position and velocity of the macroparticles \cite{Dawson-RMP-1983, Birsdall-PIC-1985, Hockney-PIC-1988}.

Typically in the PIC algorithm, Maxwell's and motion equations are solved utilising explicit time discretisation in a staggered fashion, with particles assumed to be frozen when updating the electromagnetic field and fields considered static when pushing particles~\cite{Lapenta-JCP-2012}. This methodology introduces a series of stability constraints which require that the smallest scale involved in the problem is resolved. As a consequence, using these algorithms for modelling multiscale plasma scenarios becomes very challenging even considering the large amount of computational resources available today. On the other hand, a fully implicit approach with particle and field information updated together by employing non-linear iterative schemes would relax the conditions for stability. However, the higher complexity of the algorithm, which requires the solution of a set of non-linearly coupled equations, increases its computational cost and could lead to convergence issues, thus making this solution impractical~\cite{Markidis-JCP-2011}. A good compromise is represented by semi-implicit PIC schemes. This type of algorithm maintains the benefits of an implicit discretisation of the relevant equations in time and the coupling between fields and particles is retained, though in an approximate form through linearisation. As a consequence, numerical limitations typical of the explicit method are relaxed and the absence of non-linear iterations reduces the computational cost per time step of the fully implicit PIC algorithm~\cite{Lapenta-JCP-2012, Lapenta_exascale, Gonzalez-CPC-2018, Lapenta_book}. Thanks to these features, the semi-implicit PIC algorithm has been used over the years to successfully model the coupling between small and large scales in magnetic reconnection~\cite{Lapenta_NAT_2015, Innocenti_ApJL_2015}, turbulence~\cite{franci2022anisotropic}, instabilities~\cite{micera}, and plasma confinement for fusion energy~\cite{Park_PRX_2015}.

Previously, we have implemented the semi-implicit PIC algorithm called the energy conserving semi-implicit method~\cite{Lapenta-JCP-2017, Lapenta-JPP-2017} in the massively parallel code ECsim~\cite{Gonzalez-CPC-2018, Gonzalez-CPC-2019}. The peculiarity of the scheme and the code is that both conserve energy down to machine precision contrarily to other semi-implicit schemes, such as the implicit moment method~\cite{Brackbill-1982} or the direct implicit method~\cite{Langdon-JCP-1983}. This eliminates the artificial numerical cooling typical of these algorithms, which is usually controlled in simulations by selecting sufficiently small spatial and temporal steps, thus practically increasing the computational cost of simulations. In this sense, ECsim represents a step forward towards real multiscale plasma simulations.

The code is written in C/C++ and parallelised with the Message Passing Interface (MPI\footnote{MPI: \url{https://www.mpi-forum.org/}, accessed November 2025.}) paradigm. A MPI cartesian topology is used to split the physical domain among the processes. Each process handles the field and particles on its subdomain. The code utilises the parallel libraries HDF5\footnote{HDF5: \url{https://www.hdfgroup.org/solutions/hdf5/}, accessed November 2025.} to perform data I/O and it is built using CMake\footnote{CMake: \url{https://cmake.org/}, accessed November 2025.}. A version of the code implementing OpenMP\footnote{OpenMP: \url{https://www.openmp.org/}, accessed November 2025.} shared memory parallelism on top of MPI also exists.

In this work, we describe our effort to prepare ECsim for exascale supercomputers. Normally, heterogeneity is a key characteristics of these machines, with computing nodes being equipped with accelerators, typically Graphical Processing Units (GPUs). To tackle this challenge and exploit the high level parallelism offered by these devices, we decided to adopt a directive-based approach using OpenACC\footnote{OpenACC: \url{https://www.openacc.org/}, accessed November 2025.}. OpenACC allowed us to maintain the original structure of the code without introducing intrusive changes and, at the same time, reaching good performance. It could be argued that this choice currently penalises us in terms of portability and that we could achieve the same results by using the OpenMP offloading model. However, based on our knowledge, at this stage OpenACC appears to be more stable and mature with respect to OpenMP offloading and with more support from a compiler perspective, thus guaranteeing better performance given the same effort~\cite{Stack_2021}. Furthermore two out of three European pre-exascale machines are equipped with NVIDIA GPUs and the first European exascale supercomputer, Jupiter\footnote{Jupiter: \url{https://www.fz-juelich.de/en/jsc/jupiter}, accessed November 2025.} (J\"ulich Supercomputing Centre, Germany), features NVIDIA GPUs. ECsim will be able to take full advantage of all these resources. Thus, the main contributions of this paper are:
\begin{itemize}
\item We accelerated the semi-implicit PIC code ECsim using OpenACC, requiring only minimal modifications to the original code base;
\item We validated the accelerated implementation against the reference CPU version and carried out performance benchmarks in terms of both time-to-solution and energy-to-solution;
\item We compared the execution time of the accelerated kernels across different NVIDIA GPU models and analysed the scaling behaviour of the accelerated ECsim on the pre-exascale system Leonardo (CINECA, Italy)~\cite{leonardo_turisini}.
\end{itemize}

The paper is organised as follows. Section~\ref{sec:code_structure} overviews the ECsim algorithm. Section~\ref{sec:porting} outlines the strategy adopted to introduce GPU support in the code, reports the validation tests performed to ensure correctness, and discusses performance benchmarks comparing the accelerated implementation with the reference CPU version. Experiments on different GPU architectures and scaling tests on the Leonardo supercomputer are presented in Section~\ref{sec:performance}. Finally, Section~\ref{sec:conclusion} summarises the main findings and concludes the paper. 
\section{Code structure} \label{sec:code_structure}
Three main blocks form the core of the code: 1) moment gathering, 2) field  solver, and 3) particle mover.

In the moment gathering, particle information is deposited on the grid to compute the current and the so-called mass matrices~\cite{Burgess-JCD-1992} later needed by the field solver.
In particular, the hatted current $\hat{\mathbf{J}}$ for each plasma species $s$ is calculated according to the formula:
\begin{equation}
\hat{\mathbf{J}}_{sN} = \frac{1}{V_N} \sum_p q_p \hat{\mathbf{v}}_p W_{pN}. \label{eq:explicit_cuurent}
\end{equation}
In Equation~\eqref{eq:explicit_cuurent}, $N$ denotes the set of closest grid nodes to which each particle contributes, with each particle influencing two nodes in each spatial direction (8 nodes in total), $V_N$ is the control volume of the node $N$, $W_{pN}$ the linear interpolation function between the particle $p$ of species $s$ and the node $N$, $q_p$ the charge of the particle $p$, and $\hat{\mathbf{v}}_p = \alpha_p^n \cdot \mathbf{v}_p^n$ the particle velocity $\mathbf{v}_p^n$ rotated by the matrix $\alpha_p^n$, both evaluated at time $n$.

The three-by-three rotation matrix is given by:
\begin{equation}
{\alpha}_p^n =  \frac{1}{1+(\beta_p B_p^{n})^2} \left( \mathbb{I} - \beta_p \mathbb{I} \times \mathbf{B}_p^n +\beta_p^2 \mathbf{B}_p^n \mathbf{B}_p^n \right),
\end{equation}
where $\mathbb{I}$ is the identity matrix, $\mathbf{B}_p^n$ the magnetic field at the position of the particle $p$ at time $n$, $\beta_p =  q_p \Delta t/(2 m_p)$, $m_p$ the mass of the particle, and $\Delta t$ the temporal step chosen to discretise time in the simulation.  

The calculation of the three-by-three mass matrix is more elaborated. Each $ij$ component can be found as: 
\begin{equation}
M_{NN^\prime}^{ij} = \sum_s \sum_p \frac{\beta_p}{V_N} q_p {\alpha}^{ij,n}_p W_{pN^\prime} W_{pN}, \label{mass_matrix}
\end{equation}
where $N^\prime$ indicates the neighbouring nodes to each node $N$. For each node $N$, 27 $N^\prime$ can be identified (see Figure~2 in~\cite{Gonzalez-CPC-2018}) and therefore 27 mass matrices should be calculated. Due to symmetry, $M_{NN^\prime} = M_{N^\prime N}$, only 14 mass matrices must be computed and stored in memory in practice.


In the field solver, Faraday and Amp\`ere's equations are solved to calculate the electric and magnetic fields at the next time step. The equations are discretised in time according to the $\theta$ scheme \cite{Brackbill-1982} and in space employing the technique used in the code iPIC3D~\cite{Markidis-2010, williams2024characterizingperformanceimplicitmassively}:
\begin{align}
\frac{1}{c \theta \Delta t} \mathbf{B}_C^{n+\theta} 
+ \nabla_C \times \mathbf{E}_N^{n+\theta} 
= \frac{1}{c \theta \Delta t} \mathbf{B}_C^n, \label{eq:maxwell1}\\
-\frac{1}{c \theta \Delta t} \mathbf{E}_N^{n+\theta} 
+ \nabla_N \times \mathbf{B}_C^{n+\theta} 
- \frac{4\pi}{c} \sum_{N'} M_{N N'} \mathbf{E}^{n+\theta}_p
=  \nonumber \\
= -\frac{1}{c \theta \Delta t} \mathbf{E}_N^n 
+ \frac{4\pi}{c} \hat{\mathbf{J}}_N.
\label{eq:maxwell2}
\end{align} 
Here, $c$ is the speed of light in vacuum, $\hat{\mathbf{J}}_N$ is obtained by summing $\hat{\mathbf{J}}_{sN}$ over all plasma species $s$, and the subscript $C$ denotes quantities evaluated at the cell centres, as opposed to those evaluated at the grid nodes $N$. 
The operator $\nabla$ is discretised using finite difference.
Usually, $\theta = 1/2$ is employed; this ensures energy conservation.
The system composed of Equations~\eqref{eq:maxwell1} and~\eqref{eq:maxwell2}  is solved to find $\mathbf{E}^{n+\theta}$ and $\mathbf{B}^{n+\theta}$ in an iterative way. The generalised minimal residual (GMRES) method available through the PETSc library\footnote{PETSc: \url{https://petsc.org/release/}, accessed November 2025.} is utilised. Using $\mathbf{E}^{n+\theta}$ and $\mathbf{B}^{n+\theta}$, the electric and magnetic field values at time $n+1$ are computed as $\mathbf{E}^{n+1} = (\theta-1)/\theta \mathbf{E}^{n} + 1/\theta \mathbf{E}^{n+\theta}$ and $\mathbf{B}^{n+1} = (\theta-1)/\theta \mathbf{B}^{n} + 1/\theta \mathbf{B}^{n+\theta}$.

The electric field at the iteration $n+\theta$ and the magnetic field at iteration $n$ are then used in the particle mover to update each particle velocity and position prior to interpolation from the grid point to the particle. The particle pusher is designed combining the DIM D1~\cite{Hewett1987} and the theta \cite{Brackbill-1982} schemes. First, the particle velocity at $n+1$ is computed according to the following steps:
\begin{eqnarray}
\bar{\mathbf{v}}_p &=& \alpha_p^n \cdot \mathbf{v}_p^n + \beta_p \alpha_p^n \cdot \mathbf{E}_p^{n+\theta}(\mathbf{x}_p^{n-1/2}), \\
\mathbf{v}_p^{n+1} &=& 2 \bar{\mathbf{v}}_p - \mathbf{v}_p^n,
\label{eq:vupdate}
\end{eqnarray}
and, finally, the particle is moved to a new position:
\begin{equation}
\mathbf{x}^{n+1/2} = \mathbf{x}^{n-1/2} + \Delta t \mathbf{v}^{n+1}.
\label{eq:xupdate}
\end{equation}
By choosing $\theta=1/2$, the particle mover is second order accurate in time and energy conservation is ensured~\cite{Lapenta-JCP-2017}.

These three steps are then repeated at each new time iteration. 
\section{Strategy for adding GPU support} \label{sec:porting}
\begin{figure}
\centering
\includegraphics[width=0.3\linewidth]{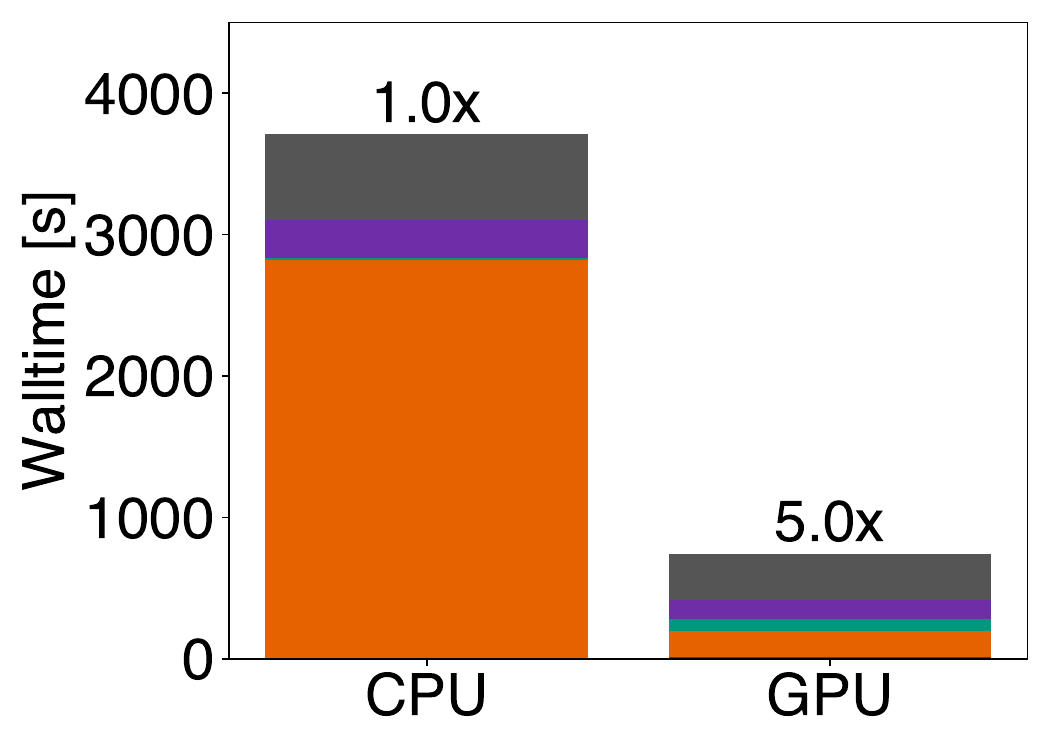}
\caption{Walltime and speedup for a typical 2D simulation using the CPU reference (left bar) and OpenACC-accelerated (right bar) ECsim on one node of Leonardo Booster. Colours denote code sections: blue initialisation, orange moment gathering, green field solver, purple particle mover, and gray I/O. The pure CPU simulation was ran with 32 MPI ranks, the heterogeneous simulation was performed with four GPUs and four MPI tasks.} \label{fig:CPU_GPU}
\end{figure}

As a first step toward accelerating ECsim, we profiled the original CPU version of ECsim on the Booster partition of the Leonardo supercomputer; this serves as the reference implementation for the rest of the text. Each compute node in this system is equipped with an Intel Xeon Platinum 8358 CPU (codename: Ice Lake), providing 32 physical cores per node, and four NVIDIA A100 GPUs, which remained idle during the CPU tests~\cite{leonardo_turisini}.
This reference version was compiled with the NVIDIA HPC SDK 23.11\footnote{NVIDIA HPC SDK 23.11: \url{https://developer.nvidia.com/nvidia-hpc-sdk-2311-downloads}, accessed November 2025} suite of compilers, Open MPI 4.1.6\footnote{Open MPI 4.1.6: \url{https://www.open-mpi.org/software/ompi/v4.1/}, accessed November 2025.}, HDF5 1.14.3, OpenBLAS 0.3.24\footnote{OpenBLAS: \url{http://www.openmathlib.org/OpenBLAS/}, accessed November 2025.}, PETSc 3.20.1, and H5HUT 2.0.0\footnote{H5HUT: \url{https://gitea.psi.ch/h5hut_new/src}, accessed November 2025.} with \verb|-O3| flag.
Figure~\ref{fig:CPU_GPU} shows the result of this exercise for a two-dimensional simulation of the current filamentation instability~\cite{CFI} using a grid of $256 \times 128$ cells, 4 plasma species, and $60 \times 60$ particles per cell per species, for a total of $ \approx4.72 \times 10^8$ particles pushed for 100 time steps. The high number of computational particles is required to decrease the numerical noise and improve the correctness and accuracy of the results~\cite{Birsdall-PIC-1985} and is a crucial parameter in ECsim, which relies solely on linear interpolation and does not implement spatial smoothing, as the latter would compromise energy conservation. We note that a recent work has proposed a smoothing technique that preserves energy conservation~\cite{Lapenta_smooth}; however, this approach is not used in our current implementation of ECsim.
The simulation was performed using $32$ MPI tasks equivalent to a full node. 
We observe that, although the grid is slightly smaller than what is typically used in production runs, the workload per core remains representative of production conditions. The number of I/O operations also closely matches that of a standard production run.

As shown by Figure~\ref{fig:CPU_GPU}, most of the time is spent by the code in the moment gathering portion ($\approx 76 \%$ of the total simulation time). This is followed by I/O and the particle mover. The I/O part accounts not only for writing to disk operations, but also for the calculation of quantities such as plasma density, current, pressure, and heat flux, necessary for diagnostic purposes rather than for temporal evolution. These computations basically entail interpolating particle to the grid. Initialisation and field solver last a negligible time with respect to the rest of the code. Thus the CPU profiling provides a clear indication of the kernels for which GPU offloading could be the most beneficial: particle mover, moment gathering, and diagnostics calculations in the I/O. Our primary focus of this work is indeed accelerating these kernels to enhance performance of ECsim, by reducing the overall time-to-solution.

We started by accelerating the particle mover, which is composed of three main functions: \verb|updataVelocity|, \verb|updatePosition|, and \verb|fixPosition|.
In the first two functions the velocity of each particle at $t^{n+1}$ and their position at $t^{n+1/2}$ are computed, respectively, employing Equations~\eqref{eq:vupdate} and~\eqref{eq:xupdate}, which use previously stored values of the electric and magnetic fields at the position of the particles. 
The third function is only called if the simulation is one or two-dimensional and set the particle position along the same line or on the same plane.
This step is necessary to ensure the correctness of the mass matrix calculations in case of reduced simulation dimensions.

The underlying structure of all these functions is similar, as each of them contains a \verb|for| loop over all particles. Since the computations exhibit no data dependencies, acceleration was achieved simply by adding the \verb|#pragma acc parallel loop| directive before the loops. As a first step, we allowed the compiler to manage memory transfers automatically and therefore did not include any explicit clauses for moving data between the host and the device.

The \verb|updateVelocity| function makes use of the field information, which was stored as three-dimensional pointers of size $N_x \times N_y \times N_z$, where $N_x$, $N_y$, and $N_z$ represent the number of grid points per MPI rank along the $x$, $y$, and $z$ directions, respectively. Before introducing the OpenACC pragmas, these three-dimensional pointers were flattened into one-dimensional arrays to improve data locality and assist the compiler in optimisation.  

We then tackled the moment gathering. The core routine in this case is \verb|computeMoments|, where $\hat{\mathbf{J}}_{sN}$ and $M_{NN^\prime}^{ij}$ are computed.
Each component of the hatted current $(\hat{{J}}_{sN,x},\hat{{J}}_{sN,y},\hat{{J}}_{sN,z})$ is represented in memory as a four-dimensional array of size $N_s \times N_x \times N_y \times N_z$ with $N_s$ indicating the number of plasma species considered. 
The nine components of the mass matrix, i.e. $M_{NN^\prime}^{xx}$, $M_{NN^\prime}^{xy}$, $M_{NN^\prime}^{xz}$, $M_{NN^\prime}^{yx}$, $M_{NN^\prime}^{yy}$, $M_{NN^\prime}^{yz}$, $M_{NN^\prime}^{zx}$, $M_{NN^\prime}^{zy}$, $M_{NN^\prime}^{zz}$, are each also stored in a four-dimensional array with size $14 \times N_x \times N_y \times N_z$ with no distinction between plasma species. As described in Section~\ref{sec:code_structure}, for each node $N$, 14 mass matrices must be computed.

The function \verb|computeMoments| 
contains a \verb|for| loop over all particles.
Within this loop, the grid node with the highest index to which each particle contributes is first identified in every spatial direction.
Because linear interpolation is used, each particle interacts with two grid points per direction, denoted $N^{-}$ and $N^{+}$.
Once $N^{+}$ is known, the corresponding lower index $N^{-}$ is simply obtained by subtracting one.
Overall, this yields $2^3 = 8$ grid points that receive contributions from a single particle.
The interpolation weights mapping the particle position to these surrounding nodes, $W_{pN}$, are then computed and stored in a $2 \times 2 \times 2$ weight array.

The weight array is used within the \verb|for| loop over the particles to calculate the value of the magnetic field at each particle position.
Originally this was done through three nested mini-loops along the dimensions of the grid with index being either 0 or 1 and mapping to either $N^{-}$ or $N^{+}$ in each direction.
These loops were manually unrolled to exploit parallelism and improve memory coalescing on the GPU, thereby reducing loop overhead and increasing computational throughput.
To aid with this both the weight and magnetic field three-dimensional arrays were flattened.

After computing the value of the magnetic field at the particle position, a rotation matrix is computed for each particle. 
In the reference CPU code, the rotation matrix is stored in a $3 \times 3$ array with elements $\alpha^{ij}$. Here, with the intent of improving data locality, we eliminated the bi-dimensional array and defined nine scalars, one for each element of the original matrix.

The current $\hat{\mathbf{J}}_{sN}$ is then computed by summing the contribution of each particle, namely $q_p \alpha_p^n \cdot \mathbf{v}_p^n$, to the neighbouring grid points.
In the original CPU version of the code this was done component by component $(\hat{{J}}_{sN,x},\hat{{J}}_{sN,y},\hat{{J}}_{sN,z})$ through again three nested mini-loops along the dimension of the grid with index being either 0 or 1.
For the same reasons listed above, we decided to unroll manually these mini-loops.

The routine proceeds by computing the 14 mass matrices, each with nine components.
In this case we made only minimal changes to the instructions used in the CPU code. Originally the computation of the mass matrix components involved six nested loops: three loops over the spatial dimensions with indexes varying between 0 and 1 to identify the closest nodes to the particle in each direction, one loop from 0 to 14 for the neighbour nodes $N'$, and finally two loops to effectively compute the contribution of $p$ to the three-by-three $(i,j)$ component of each mass matrix. These latter values were then accumulated to $M_{NN^{\prime}}$, with \verb|if| conditions used to select the nodes to which each particle has to accumulate.

First, we observed that the access pattern to the four-dimensional arrays storing the mass matrices was suboptimal in terms of memory locality.
Consequently, we modified the access order to maintain the leftmost dimension (loop over $N^{\prime}$) as the outermost loop.
This implied some further modifications.
We introduced a \verb|switch| construct to identify the correct neighbour to the node $N$ among the 14 possibilities.
We maintained three \verb|for| mini-loops over the grid dimensions with indexes varying between 0 and 1. This, together with \verb|if| conditions to exclude nodes too far from the particle, allowed us to identify both the relevant nodes and the amount to deposit there, e.g. $q_p \beta \alpha^{ij} W_{pN'}W_{pN}$. The overall four \verb|for| loops serve to sum also the different contributions to the nine components of each mass matrix. 



To accelerate \verb|computeMoments|, after the modifications we have mentioned, we added the directive \verb|#pragma acc parallel loop| before the \verb|for| loop over the particles.
Additionally, to avoid race conditions caused by concurrent threads accessing the same memory location, we had to resort to \verb|#pragma acc atomic update| before summing the contributions for $\hat{\mathbf{J}}_{sN}$ and $M_{NN^\prime}^{ij}$.

We remark that the routine still contains three relatively small inner loops. Full loop unrolling was not applied, as it would increase register pressure and consequently degrade execution performance.  

Diagnostics for I/O are computed through two routines, \verb|computeCharge| and \verb|computeCurrent|, which evaluate the charge density, charge current, and pressure tensor. The diagnostic charge current differs from $\hat{\mathbf{J}}$, as it excludes the effect of velocity rotation. The structure of these routines closely resembles that of \verb|computeMoments|: each contains a \verb|for| loop over the particles and a summation that accumulates their individual contributions. To accelerate these computations, the loops were parallelised using \verb|#pragma acc parallel loop|, and the summations were made thread-safe through the use of \verb|#pragma acc atomic update|.

We note that the use of \texttt{\#pragma acc atomic update} could potentially be avoided by sorting the particles and employing a reduction clause for the summation. However, when we attempted to implement this approach, compiler limitations arose with array based reductions, and we decided not to pursue this approach further.

Performance analysis using NVIDIA Nsight Systems revealed substantial data transfers between the host and device, accompanied by numerous page faults during GPU kernel execution. To address this, we chose to continue using managed memory while leveraging the CUDA memory management function \verb|cudaMemPrefetchAsync| to efficiently manage data migration between the host and device. Specifically, \texttt{cudaMemPrefetchAsync(x, sizeof(double) * n, 0, 0)} was used to transfer an array \verb|x| of \verb|n| double elements to the GPU before its use, while \texttt{cudaMemPrefetchAsync(x, sizeof(double) * n, cudaCpuDeviceId, 0)} was employed to migrate the same array back from the device to the host. This approach improves performance while circumventing the intricate issues associated with deep-copying complex data structures.

\begin{figure}
\centering
\includegraphics[width=0.6\linewidth]{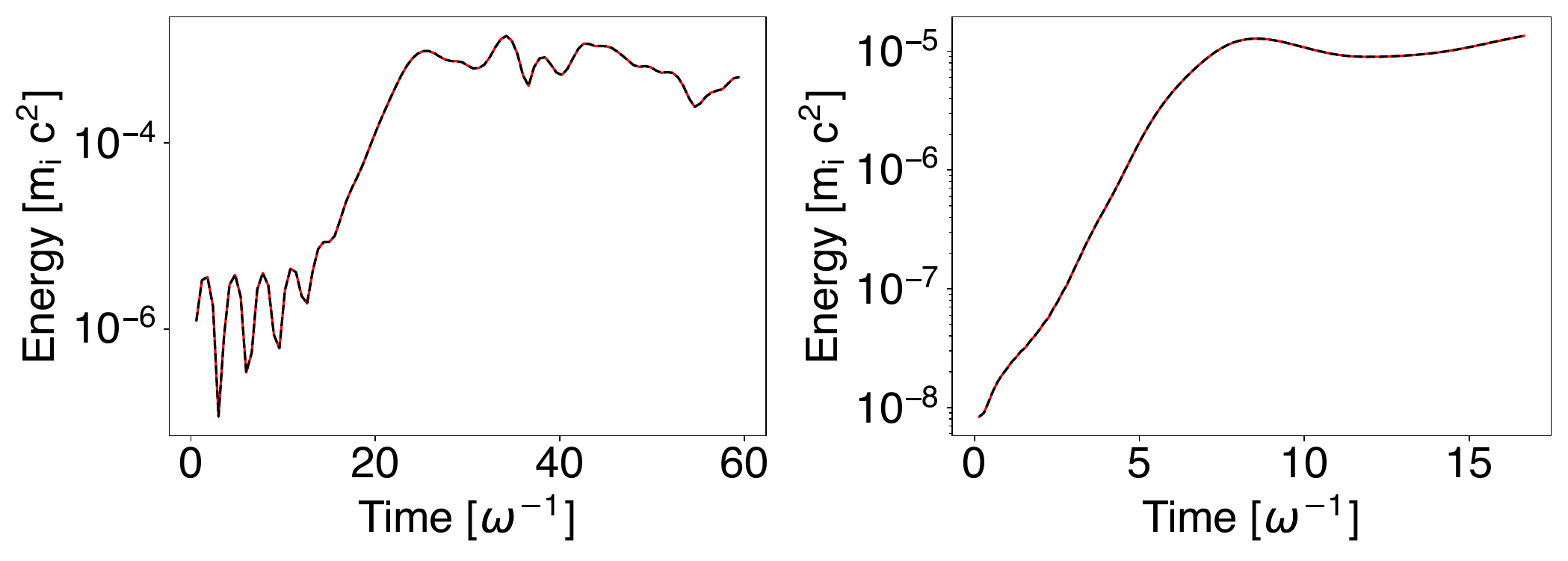}
\caption{Time evolution of the electric field energy (left panel) and magnetic field energy (right panel) from simulations of the two-stream instability (left panel) and current filamentation instability (right panel), respectively. Red solid lines show results from the CPU reference version of ECsim, while black dashed lines correspond to the accelerated version of the code. All values are normalised to reference quantities.} \label{fig:physics_CPU_GPU}
\end{figure}

Before evaluating the performance of the accelerated code, we first validated its accuracy by comparing the results against those obtained with the CPU reference version. Specifically, we performed simulations on the Leonardo Booster partition of well-known one- and two-dimensional plasma instabilities using both code variants. The accelerated ECsim was compiled with the same software stack as the reference version, but with the following additional compiler flags:
\texttt{-cuda -acc -gpu=cc80,managed,lineinfo,cuda12.3 -Minfo=accel}, along with \verb|-O3| for optimisation.
Both one- and two-dimensional simulations were executed using four MPI tasks on the CPU, and four MPI tasks, each bound to a distinct NVIDIA A100 GPU, when running with accelerator support.
The Open MPI flags \verb|--bind-to core| and \verb|--map-by ppr:4:node:PE=8| were used in both cases.
Figure~\ref{fig:physics_CPU_GPU} shows the temporal evolution of the electric and magnetic field energies for the two-stream instability~\cite{two_stream} (left panel) and the current-filamentation instability~\cite{CFI} (right right), standard reference cases for electrostatic and electromagnetic dynamics, respectively. The results display the characteristic exponential growth typical of these phenomena. The curves produced by the GPU-accelerated version of ECsim overlap perfectly with those from the reference implementation, confirming the correctness and numerical consistency of the accelerated code.

Once the correctness of the code had been established, we measured its performance by repeating the same simulation shown in Figure~\ref{fig:CPU_GPU} using the four GPUs available on a node of the Booster partition of Leonardo each associated to one MPI rank. For the performance evaluation, a full node was allocated exclusively. The same Open MPI flags and GPU binding techniques used for Figure~\ref{fig:physics_CPU_GPU} were employed. Execution times are reported in Figure~\ref{fig:CPU_GPU}. 
The GPU-accelerated version of the code completed the simulation in $\approx 787 \, \mathrm{s}$, achieving an overall speedup of $5\times$ relative to the reference CPU implementation, which utilised all the 32 cores available on one node. The moment-gathering block now accounts for only $23 \%$ of the execution time, with a duration comparable to the other code blocks. We note that this overall $5\times$ speedup was achieved with partial offloading of the code. Functions that remain on the CPU and are parallelised solely via MPI, such as initialisation and field calculation, are approximately eight times slower in our experiment due to the lower number of MPI tasks (four vs 32) used in the offloaded version. In contrast, the moment gathering block is accelerated by a factor of $15\times$, while the field solver and I/O routines are roughly twice as fast.

\begin{figure}
\centering
\includegraphics[width=0.4\linewidth]{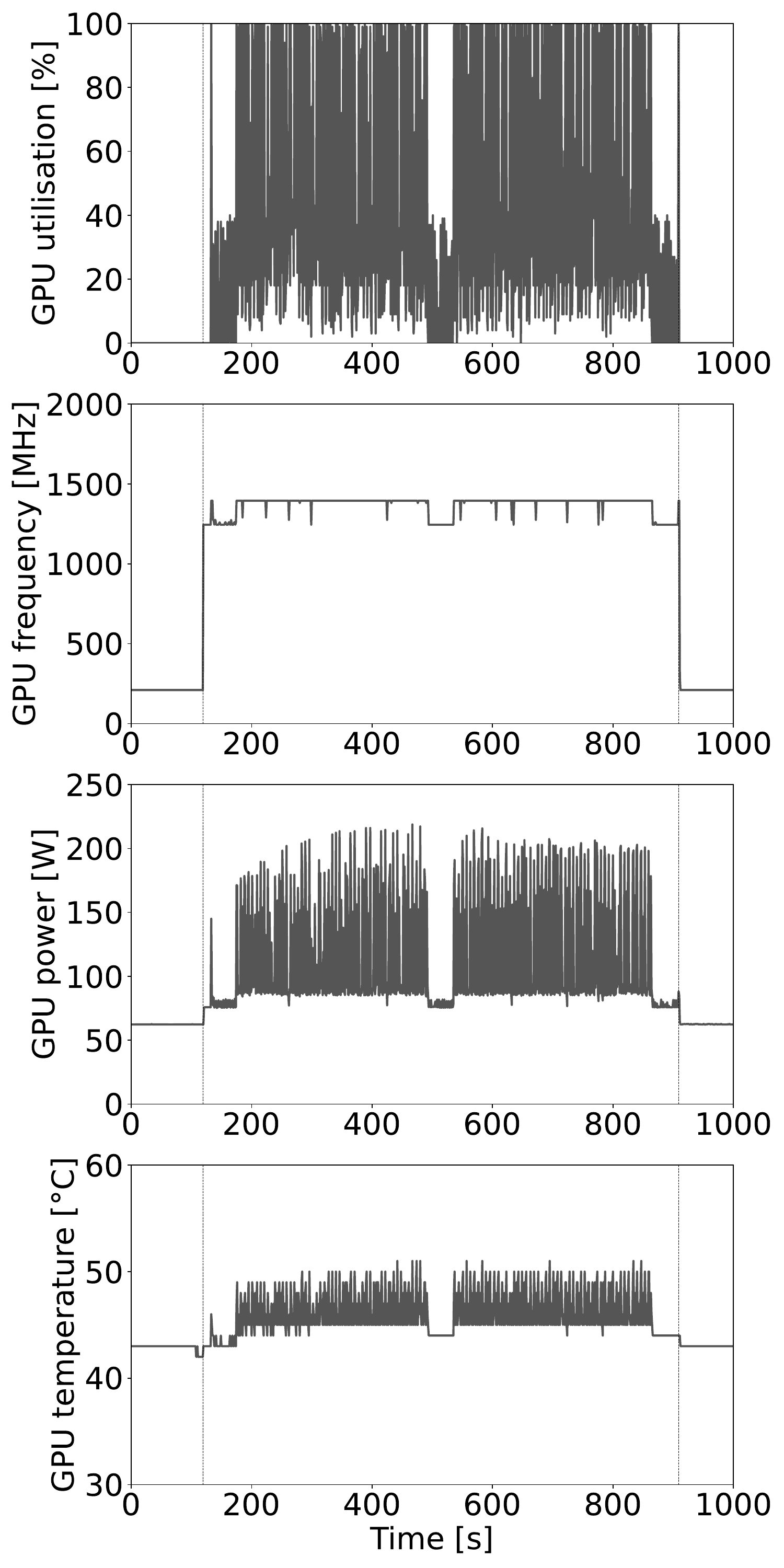}
\caption{GPU utilisation (first row), frequency (second row), power (third row), and temperature (fourth row) during an ECsim run. The same use case as in Figure~\ref{fig:CPU_GPU} was employed. Four GPUs were utilised, but only the values for one representative device are shown here due to the strong similarity across all GPUs.}\label{fig:nvsmi}
\end{figure}

Using the NVIDIA System Management Interface (\verb|nvidia-smi|), we monitored the behaviour of the four GPUs available on a Leonardo Booster node during a series of ten simulations based on the run shown in Figure~\ref{fig:CPU_GPU}. In particular, \verb|nvidia-smi| was run in the background in user space throughout each simulation, and its output was logged to a text file at a frequency of $\approx 1 \,\mathrm{Hz}$, following the methodology in~\cite{Amati, Almerol}.
Figure~\ref{fig:nvsmi} shows the time evolution of the GPU utilisation fraction, streaming multiprocessor (SM) frequency, power, and temperature for one of the four GPUs during a representative run. The other GPUs, not shown here, exhibit the same behaviour, and these trends are consistently observed across all ten simulations.

Each simulation was preceded and followed by an interval of $120 \, \mathrm{s}$ \verb|sleep| to ensure that the GPUs returned to idle conditions between consecutive runs. During these periods, GPU utilisation remained at $0$, and the clock frequency, power, and temperature stabilised at idle values, namely, a frequency of $210 \, \mathrm{MHz}$, a power consumption of $\approx 60 \, \mathrm{W}$, and a temperature oscillating between $42$ and $43 \, ^{\circ}\mathrm{C}$. The beginning and end of the ECsim simulation are indicated by vertical black lines in the figure.

At the start of the simulation, a short initialisation phase lasting less than $2 \, \mathrm{s}$ occurs, during which the GPU remains in an idle state. Immediately afterwards, results at $t = 0$ are dumped to disk. This operation produces a peak in GPU utilisation and in all related parameters because, as discussed in Section~\ref{sec:porting}, the calculations of plasma density, current, and pressure are performed on the GPU.

During the subsequent write-to-disk phase, GPU metrics exhibit a local minimum, although still above idle levels. This I/O step is longer than usual because RAW particle data are saved at this point; given the number of particles and despite parallelisation, this operation requires $\approx 50 \, \mathrm{s}$. While grid data are saved at every cycle and require only a few seconds, RAW particle data are saved three times during the simulation, and these events are clearly recognisable in Figure~\ref{fig:nvsmi}. During these heavy I/O operations, only the CPU is active; consequently, GPU-related parameters temporarily drop to local minima.

Outside these I/O phases, GPU utilisation oscillates between $100 \%$, when GPU kernels are executed, and very low values when the time step enters code sections that are not offloaded. Conversely, the SM clock frequency shows limited variation. Except during the particle dump to disk, it remains at $1395 \, \mathrm{MHz}$, slightly below the maximum frequency of $1410 \, \mathrm{MHz}$. GPU power and temperature follow the utilisation pattern, reaching peaks of $\approx 210 \, \mathrm{W}$ and $\approx 50 \, ^{\circ}\mathrm{C}$, respectively.

From these measurements several considerations can be drawn. First, the GPU is efficiently saturated during the phases where kernels are launched, as indicated by utilisation spikes to $100 \%$ and the corresponding increases in power and temperature. Second, the long plateaus at low utilisation highlight that the overall performance is partially constrained by CPU-side work and, in particular, by I/O. The pronounced local minima during RAW particle dumps indicate that the GPU remains mostly idle during heavy disk operations. Finally, the nearly constant SM frequency during compute phases indicates that the GPU operates close to its intended performance envelope, without significant thermal throttling or power capping. This confirms that the observed performance fluctuations are driven by algorithmic and I/O patterns rather than hardware limitations.

With the goal of understanding the energy consumption of a typical simulation, we also monitored the energy absorbed by the CPU. To obtain meaningful statistics, we launched ten simulations with the reference CPU-only code using 32 MPI tasks and ten simulations with the accelerated code using 4 MPI tasks and 4 GPUs. Simulations were not constrained to the same physical node, so results reflect a representative sample across multiple nodes. The physical problem solved corresponds to the same use case shown in Figure~\ref{fig:CPU_GPU}. CPU energy consumption was measured using \verb|perf stat -a -e| in combination with a one-second \verb|sleep| interval. On Leonardo Booster, unlike on other supercomputers, this command does not require \verb|sudo| privileges and can therefore be used directly in user space. As with GPU power measurements, CPU energy samples were collected in the background while the simulations were running and output to a text file. We note that this analysis focuses solely on CPU and GPU; energy contributions from RAM, disks, network, and other node components are neglected.

\begin{figure}
\centering
\includegraphics[width=0.6\linewidth]{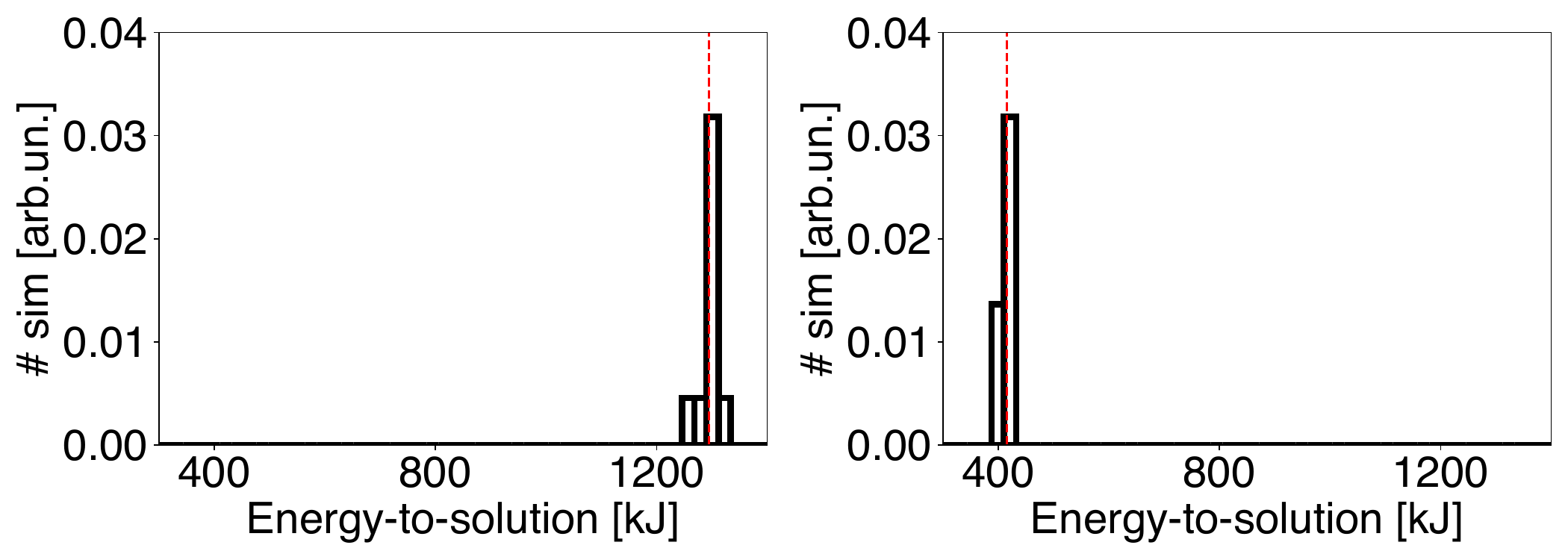}
\caption{Distribution of energy values across ten simulations for runs on CPU only (left panel) and on GPU (right panel). The red dashed lines represent the average energy-to-solution. Same input parameters and number of MPI tasks and GPUs as in Figure~\ref{fig:CPU_GPU} were used.}\label{fig:energy}
\end{figure}

Starting from the power measurements, the energy-to-solution for each NVIDIA A100 GPU was computed by performing a discrete integration of the recorded power samples over the simulation time (excluding the sleep intervals). The total accelerator energy was then obtained by summing the contributions from all four GPUs. The processor energy was computed analogously by summing over values collected with \verb|perf|, again considering only the simulation time. The total energy-to-solution of a single run was therefore obtained as the sum of CPU and GPU contributions. Figure~\ref{fig:energy} shows the distributions of total energy for the ten CPU-only simulations (left panel) and the ten CPU+GPU simulations (right panel). On average, the reference CPU code required $1294.74 \pm 16.72 \, \mathrm{kJ}$, whereas the accelerated code consumed $415.33 \pm 5.55 \, \mathrm{kJ}$, demonstrating a $\approx 3 \times$ improvement in energy efficiency. The absolute reduction in energy consumption stems from two complementary effects: the accelerated code both reduces time-to-solution and shifts the bulk of the computation to GPUs, which offer significantly higher performance per watt compared to CPUs for these kernels. However, the CPU contribution in the accelerated runs remains non-negligible, reflecting the fact that ECsim still performs significant CPU-side operations, including I/O, domain decomposition updates, and other tasks not offloaded to the GPU. The ratio between CPU and GPU energy suggests that further optimisations, such as optimising heavy I/O phases, improving CPU–GPU overlap, or offloading additional routines, could yield even larger energy savings.

\section{Performance evaluation} \label{sec:performance}

We compared the performance of the GPU-enabled ECsim on different generation of NVIDIA GPUs. In particular we compared V100, A100, H100, and GH200. For completness, we report here also the type of hosting system in the case of the first three accelerators: IBM Power 9, Intel Cascade Lake, and AMD Genoa. 

For this comparison, we performed simulations using a two-dimensional grid with $128 \times 128$ cells, each containing 6400 computational particles. The simulations comprised 596 temporal iterations. All the simulations were performed using 1 MPI task and 1 GPU.  

\begin{figure}
\centering
\includegraphics[width=0.6\linewidth]{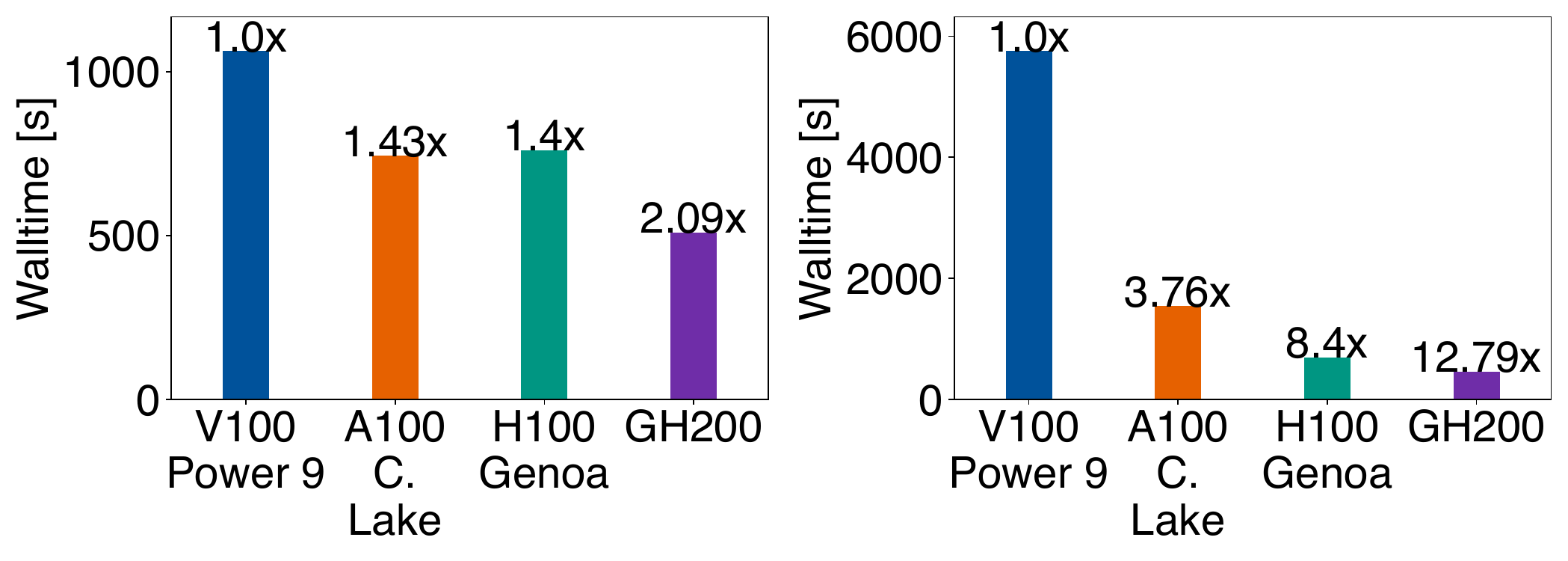}
\caption{Walltime results for the particle mover (left panel) and moment gathering (right panel) obtained on various NVIDIA GPU generations. The corresponding CPU configurations are provided for completeness. Speedup values shown above the bars are computed relative to the walltime obtained on the V100 GPU.} \label{fig:generation_GPU}
\end{figure}

Results are reported in Figure~\ref{fig:generation_GPU}, showing the execution time of the particle mover (left panel) and moment gathering (right panel) across different NVIDIA GPU generations. In both cases, execution time decreases as newer GPUs are employed.

For the particle mover, we observe a modest speedup of about $1.4 \times$ on both A100 and H100 compared to V100, which exhibits the slowest performance. Kernels in the particle mover are primarily memory-bound. Their performance depends more on memory latency and bandwidth than on the available compute capability.
Both A100 and H100 feature similar memory-bandwidth scaling for global memory operations when accessed through OpenACC-managed loops, explaining the limited improvement between them. On GH200, however, the particle mover runs about twice as fast as on the V100. This gain can be attributed to the unified CPU–GPU memory architecture of the GH200, where CPU and GPU share the same physical memory pool via NVLink-C2C interconnect. In this configuration, the prefetch operations effectively become low-overhead cache hints rather than explicit host–device transfers, eliminating PCIe data movement costs. As a result, memory accesses are more uniform and latency is significantly reduced.

The moment gathering kernel shows a much stronger performance improvement. Speedups relative to V100 reach $3.76 \times$, $8.4 \times$, and $12.79 \times$ on A100, H100, and GH200, respectively. This behaviour can be explained by a combination of factors. Newer GPUs have larger memory bandwidth, which directly benefits a memory-bound kernel like this one. Both A100 and especially H100 implement more efficient atomic instructions at the hardware level, drastically reducing contention and serialisation penalties during accumulation operations. They are equipped with larger and more efficient L2 caches, which help coalesce and buffer atomic writes, mitigating contention. Finally, as with the particle mover, GH200 shared memory eliminates the need for costly data migration, further boosting performance.

Finally, in both cases, the CUDA driver may also have contributed to the performance improvements, as more recent CUDA versions (available on newer systems) provide better optimisation of OpenACC-generated kernels.

We performed strong and weak scaling tests of the accelerated version of ECsim on the Booster partition of Leonardo. In these tests, the dynamics of a warm plasma were modelled. In such simulations, particles move between nodes and place stress on memory access, making them representative of typical production runs.

\begin{figure}
\centering
\includegraphics[width=0.5\linewidth]{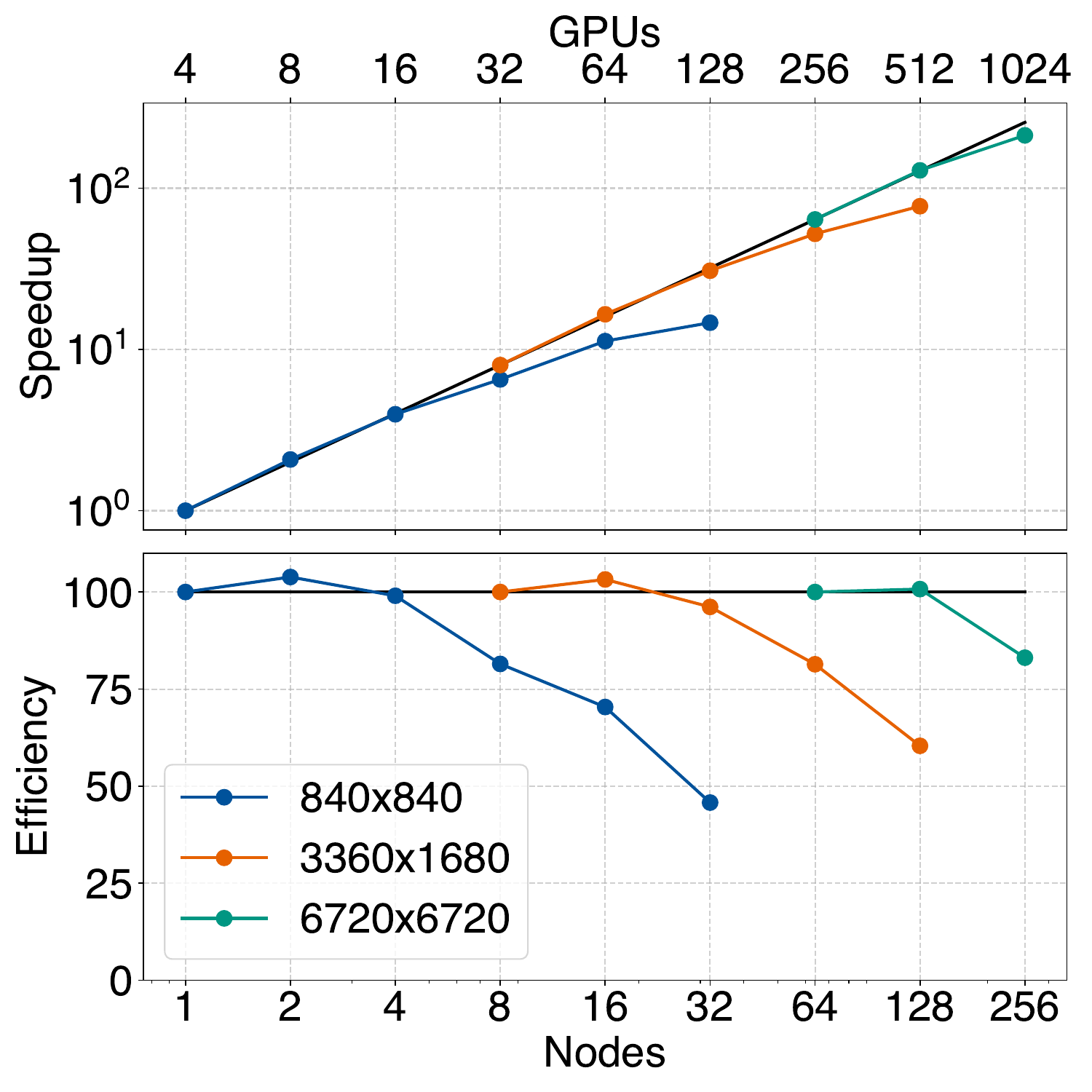}
\caption{Parallel speedup (top panel) and efficiency (bottom panel) from strong scaling tests on the Leonardo Booster partition for problem sizes of $840 \times 840$ (blue), $3360 \times 1680$ (orange), and $6720 \times 6720$ (green) cells. Black lines indicate the ideal values.} \label{fig:strong_scaling}
\end{figure}

For the strong scaling tests, three sets of simulations were carried out using grids of $840 \times 840$, $3360 \times 1680$, and $6720 \times 6720$ cells, starting with $4$, $32$, and $256$ GPUs, respectively. All simulations used approximately $3000$ particles per cell and 100 time steps. Results are reported in Table~\ref{tab:strong_scaling} and shown graphically in Figure~\ref{fig:strong_scaling}, which displays the speedup (top panel) and the efficiency (bottom panel).
These tests demonstrate that ECsim achieves near-ideal speedup for GPU counts up to 64.
For larger numbers of GPUs, the parallel efficiency gradually decreases, likely as a consequence of increased communication overhead and memory access costs.
A similar behaviour has been observed for the iPIC3D code~\cite{Markidis-2010}, whose underlying MPI parallelisation is closely related to that of ECsim.
Indeed, as shown in~\cite{SPACE1,SPACE2}, the time spent in MPI communications and the associated latency increase with the number of processes, causing a significant degradation of transfer efficiency when simulations employ a large number of cores.
Although this analysis refers to a pure MPI implementation, the introduction of GPU acceleration can further amplify these effects, since faster on-device computations increase the relative impact of inter-process communication.
In order to maintain a parallel efficiency above $70\%$ for 64 GPUs and beyond, it was therefore necessary to increase the grid size.
With these larger grid, we observed quasi-linear speedup from 32 to 256 GPUs, with a measured parallel efficiency of $81\%$, after which an even larger problem size had to be considered.
In the final set of simulations with a grid of $6720 \times 6720$ cells, starting from 256 GPUs, we achieved a parallel efficiency of $83\%$ at 1024 GPUs, confirming that ECsim can sustain high performance even at large scale.

\begin{table}[]
\centering
\begin{tabular}{|r|r|r|r|r|}
\hline
Nodes & GPUs & Walltime [s] & Speedup & Efficiency [\%] \\
\hline
1   & 4    & 2421.48 & 1.00   & 100.00 \\
2   & 8    & 1165.74 & 2.08   & 103.86 \\
4   & 16   & 611.30  & 3.96   & 99.03  \\
8   & 32   & 371.50  & 6.52   & 81.48  \\
16  & 64   & 215.02  & 11.26  & 70.38  \\
32  & 128  & 165.32  & 14.65  & 45.77 \\
\hline
8   & 32   & 2732.51 & 8.00   & 100.00 \\
16  & 64   & 1323.41 & 16.52  & 103.24 \\
32  & 128  & 710.40  & 30.77  & 96.16  \\
64  & 256  & 419.65  & 52.09  & 81.39  \\
128 & 512  & 282.80  & 77.30  & 60.39  \\
\hline
64  & 256  & 2978.46 & 64.00  & 100.00 \\
128 & 512  & 1478.11 & 128.96 & 100.75 \\
256 & 1024 & 896.22  & 212.70 & 83.08  \\
\hline
\end{tabular}
\caption{Strong scaling results for three problem sizes: $840 \times 840$ (top table), $3360 \times 1680$ (middle table), and $6720 \times 6720$ (bottom table) cells.} \label{tab:strong_scaling}
\end{table}

\begin{figure}
\centering
\includegraphics[width=0.5\linewidth]{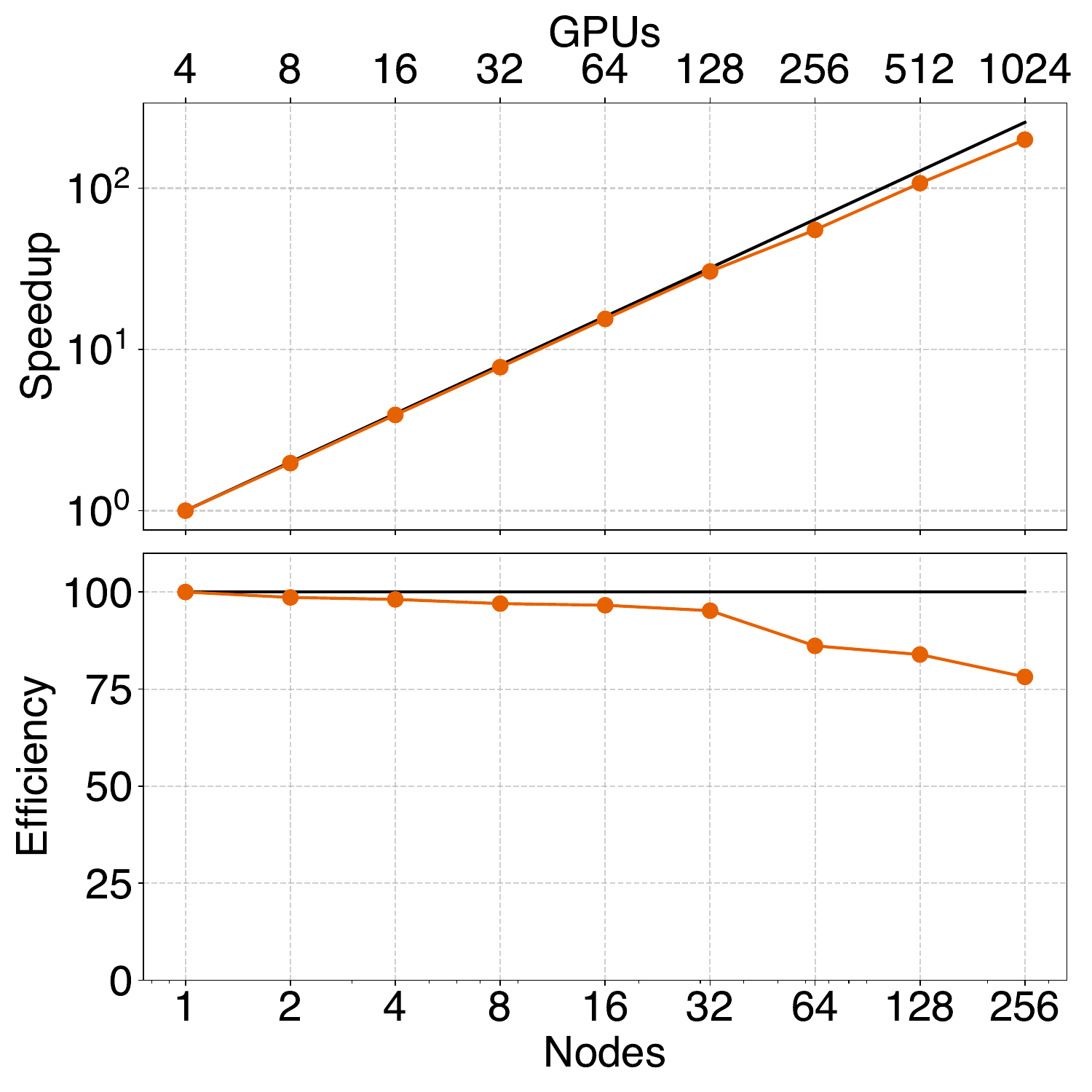}
\caption{Same as in Figure~\ref{fig:strong_scaling}, but for weak scaling tests.} \label{fig:weak_scaling}
\end{figure}

In the weak scaling tests, we started with simulations using four GPUs on a single Leonardo Booster node with a grid of $128 \times 128$ cells. The grid was then progressively increased up to $2048 \times 2048$ cells, utilising 256 Leonardo Booster nodes (1024 GPUs), which corresponds to the maximum number of nodes a single user can request on this system. Each simulation employed $200 \times 200$ particles per cell per species for a total of four plasma species, and 100 temporal steps were executed. Table~\ref{tab:weak_scaling} and Figure~\ref{fig:weak_scaling} present the results, demonstrating that ECsim exhibits excellent weak scaling. As the number of GPUs increases, the runtime remains nearly constant, indicating that communication and memory overheads grow moderately even at large scale. The measured parallel efficiency remains above $78.13\%$ for $1024$ GPUs, corresponding to a speedup that closely follows the ideal linear trend for weak scaling. These results confirm that ECsim can effectively utilise large numbers of GPUs while maintaining high computational performance, making it suitable for large-scale plasma simulations where particle motion between nodes can stress memory access and communication patterns.

\begin{table}[]
\centering
\begin{tabular}{|r|r|r|r|r|}
\hline
Nodes & GPUs & Walltime [s] & Speedup & Efficiency [\%] \\
\hline
1   & 4    & 2416.41 & 1.00   & 100.00 \\
2   & 8    & 2450.68 & 1.97   & 98.60  \\
4   & 16   & 2463.25 & 3.92   & 98.10  \\
8   & 32   & 2490.84 & 7.76   & 97.01  \\
16  & 64   & 2501.75 & 15.45  & 96.59  \\
32  & 128  & 2538.33 & 30.46  & 95.20  \\
64  & 256  & 2805.62 & 55.12  & 86.13  \\
128 & 512  & 2880.18 & 107.39 & 83.90  \\
256 & 1024 & 3092.61 & 200.03 & 78.13  \\
\hline
\end{tabular}
\caption{Weak scaling results.} \label{tab:weak_scaling}
\end{table}

\section{Summary \& Perspectives} \label{sec:conclusion}

In this work, we describe our strategy for offloading the most computationally intensive kernels of the PIC code ECsim using OpenACC directives, which allows us to maximize performance without requiring extensive code restructuring.

Our performance evaluation, in terms of both time-to-solution and energy-to-solution, using a node-to-node comparison on the Leonardo Booster partition, shows that the accelerated code achieves $5\times$ faster execution and $3\times$ higher energy efficiency compared to the CPU version.

We further evaluated the performance of the ported kernels across multiple generations of NVIDIA GPUs. The results show that they benefit significantly from the unified memory available on the GH200 Superchip, while the larger memory bandwidths of newer GPUs contribute substantially to the acceleration of memory-bound kernels in ECsim. Additionally, a kernel that relies heavily on atomic operations experiences notable speedup thanks to the more efficient hardware implementation of these instructions in modern GPU architectures.

Furthermore, we demonstrate that ECsim scales effectively across multiple GPUs. In strong scaling tests, the code achieves a parallel efficiency of $70 \%$ up to 64 GPUs (16 nodes), while in weak scaling tests it maintains a parallel efficiency of $78 \%$ up to 1024 GPUs (256 nodes).

Finally, we note that while comparisons with native multi-GPU PIC implementations or with other highly optimised open-source PIC codes would undoubtedly enrich the discussion, such comparisons are neither straightforward nor entirely fair. ECsim is intrinsically distinct from many existing PIC codes due to the numerical scheme it employs. In contrast, many highly optimised PIC codes based on pure CUDA~\cite{ipic_gpu, Chien_2020, Gordon_Bell}, Kokkos~\cite{PIC_Kokkos}, Alpaka~\cite{PIC_on_GPU}, or SYCL~\cite{PIC_SYCL} rely on different algorithmic choices, data layouts, and execution models, making direct comparisons based solely on raw performance metrics difficult to interpret. For this reason, the performance results presented in this work are primarily benchmarked against the original CPU implementation of ECsim, which provides a consistent and meaningful reference for assessing the impact of GPU acceleration.

Comparisons with other PIC codes were previously carried out during the validation phase of ECsim, most notably against iPIC3D, which shares a similar MPI parallelisation strategy. These studies focused on numerical accuracy and physical fidelity, and only partially on performance. They showed that, in one-to-one comparisons, ECsim was slightly slower than iPIC3D; however, it could achieve faster times-to-solution by exploiting larger time steps and coarser spatial resolutions while maintaining the same level of accuracy~\cite{Gonzalez-CPC-2018, Gonzalez-CPC-2019}.

While native GPU implementations of ECsim based on CUDA, Kokkos, or SYCL could in principle be developed, such approaches would require intrusive refactoring and a substantial redesign of the existing production-ready code. The primary objective of the present work is instead to enable efficient exploitation of GPU architectures while preserving the original code structure. To this end, a directive-based OpenACC approach was adopted, offering a practical compromise between performance, portability, and development effort. Nevertheless, as part of our ongoing work, a pure CUDA implementation of ECsim is currently under development and will provide valuable insight into the potential performance benefits of a fully native GPU approach.

Finally, as a future perspective, we plan to explore the use of cellular automata and high-performance libraries specifically designed for cellular automata as a means to improve performance~\cite{ref1, ref2}.
 
\section*{Acknowledgment}
\noindent E.B. and N.S. contributed equally to this work.\\
E.B. and N.S. acknowledge support from the SPACE project, funded by the European Union. This project has received funding from the European High Performance Computing Joint Undertaking (JU) and from Belgium, the Czech Republic, France, Germany, Greece, Italy, Norway, and Spain under grant agreement No.~101093441.\\
M.E.I. acknowledges support from the Deutsche Forschungsgemeinschaft (German Research Foundation), project numbers 497938371 and 544893192.\\
E.B. would like to thank Daniele Gregori (E4 Computer Engineering) for fruitful discussions.\\
This work is dedicated to the memory of Prof. Giovanni Lapenta, whose guidance and encouragement inspired this study.\\


\bibliographystyle{IEEEtran}
\bibliography{biblio}

\end{document}